\documentclass[
  twocolumn,
  prl,
  showpacs,
  amsmath,
  amssymb,
  superscriptaddress
 ]{revtex4-1}

\usepackage{color}
\usepackage{soul}
\usepackage{bm}
\usepackage{bbm}
\usepackage{graphicx}
\usepackage{amsmath}
\usepackage{hyperref}
\usepackage{todonotes}
\usepackage{epsfig}
\usepackage{psfrag}

\begin{document}

\newcommand{\cblue}{\color{blue}}
\newcommand{\cred}{\color{red}}
\newcommand{\mpar}[1]{{\marginpar{\it#1}}}
\def \be{\begin{equation}}
\def \ee{\end{equation}}

\title{
Spin-mediated particle transport in the disordered Hubbard model}

\author{Ivan V. Protopopov}
\affiliation{Department of Theoretical Physics, University of Geneva, 1211 Geneva, Switzerland  }
\affiliation{
 L.\ D.\ Landau Institute for Theoretical Physics RAS,
 119334 Moscow, Russia
}
\author{Dmitry A. Abanin}
\affiliation{Department of Theoretical Physics, University of Geneva, 1211 Geneva, Switzerland  }

\begin{abstract}
Motivated by  the recent experiments that reported signatures of many-body localization of ultracold atoms in optical lattices [M. Schreiber {\it et al.}, Science {\bf 349}, 842 (2015)], we study dynamics of highly excited states in the strongly disordered Hubbard model in one dimension. Owing to the $SU(2)$ spin symmetry, spin degrees of freedom form a delocalized thermal bath with a narrow bandwidth. The spin bath mediates slow particle transport, eventually leading to delocalization of particles. The particle hopping rate is exponentially small in $t/W$ ($t$, $W$ being hopping and disorder scales) owing to the narrow bandwidth of the spin bath. We find the optimal lenghtscale for particle hopping, and show that the particle transport rate depends strongly on the density of singly occupied sites in the initial state. The delocalization rate is zero for initial states with only doubly occupied or empty sites, suggesting that such states are truly many-body localized, and therefore the Hubbard model  may host both localized and delocalized states. Full  many-body localization can be induced by breaking spin rotational symmetry. 

\end{abstract}
\date{\today}

\maketitle

{\bf Introduction.} The phenomenon of many-body localization (MBL) has been attracting significant theoretical \cite{Anderson80,Basko06,Mirlin05,Znidaric08,PalHuse, Serbyn13-1,Huse13,Vosk13, ScardicchioLIOM,Alet-frac,Huse13L,Ponte15, Lazarides15, VasseurHotChains, PotterSymmetry, Roeck17} and experimental \cite{Bloch15, Bloch16, Monroe16, xu2018, Shahar, Choi16DTC, Rubio-Abadal2018, Bordia17, Lukin2018} interest over the past few years, see Refs.~\cite{AbaninReview,Huse-rev,ParameswaranVasseurReview2018} for recent reviews. MBL provides a mechanism of ergodicity breaking in quantum many-body systems. Ergodicity breaking has been understood as the consequence of emergent, robust integrability~\cite{Serbyn13-1,Huse13,ScardicchioLIOM} -- the property which is also responsible for the largely universal dynamical properties of MBL systems, such as logarithmic growth of entanglement entropy following a quantum quench~\cite{Znidaric08,Moore12,Serbyn13-2}, as well as power-law relaxation of local observables~\cite{,Serbyn14}. 

Recently, signatures of MBL have been observed in experiments with ultracold atoms in optical lattices~\cite{Bloch15}. The experimental system of Ref.~\cite{Bloch15} can be modeled as a fermionic Hubbard model subject to a quasi-random potential. This model is characterized by the high $SU(2)$ spin symmetry, in contrast to the less symmetric models of MBL which have been extensively studied theoretically. Recently, it has been argued that continuous non-Abelian symmetries destroy MBL in spin systems~\cite{VasseurHotChains,PotterSymmetry,Abanin16}; intuitively, this stems from the fact that such symmetries inevitably lead to degeneracies in the energy spectrum, which, in turn, induces resonances. 
Thus, it is important to understand whether the experimental system of Ref.~\cite{Bloch15} exhibits true localization, and whether the delocalization of spin degree of freedom may lead to (possibly very slow)  transport of particles. 

Motivated by experiment~\cite{Bloch15}, in this paper we study dynamics and highly excited eigenstates in the disordered one-dimensional Hubbard model:
\begin{equation}
H_0=t\sum_{\langle ij\rangle,\sigma=\uparrow,\downarrow}c^{+}_{i\sigma}c_{j\sigma}+ \sum_{i}\epsilon_i c_{i\sigma}^+c_{i\sigma}+U\sum_{i}n_{i\uparrow}n_{i\downarrow}, 	
\label{Eq:H0}
\end{equation}
where the first term represents hopping between nearest neighbor sites, the second term describes disorder potential, and the last term is the Hubbard interaction. We will assume that on-site energies are random uncorrelated variables, $\epsilon_i\in [-W;W]$. For simplicity, we focus on the limit of strong disorder, $W\gg t$. Then, the single-particle problem (with $U=0$) is in the strongly localized regime, with the localization length $\xi\sim 1/\ln(W/t)$. 
In the interacting case, recent theoretical works~\cite{Znidaric16,Prelovsek18, Zakrzewski2018} found that the spin degree of freedom in the model (\ref{Eq:H0}) remains delocalized and exhibits sub-diffusive transport~\footnote{According to Ref. \cite{Zakrzewski2018} the exponent characterising subdhiffusive transport can be very small. }, in agreement with the general arguments of Refs.~\cite{PotterSymmetry,Abanin16}.

We consider a quantum quench setup: the system is initialized at $t=0$ in a product state, where different lattice sites are singly occupied, doubly occupied, or empty, $n_i(t=0)=0,1,2$. 
A version of this setup with $n_i=0,1$ on even/odd sites was studied experimentally in Ref. \cite{Bloch15}, and the decay of such charge-density wave configuration was probed. In addition, the effect of adding a certain density of doublons, $n_i=2$, on localization was investigated.   We are interested in understanding the  dynamics of particles mediated by the coupling to the delocalized spins, and, in particular, whether/how quickly the initial density modulation decays. As we will see below, the dynamics of 'charge' degrees of freedom depends strongly on the initial density of singly occupied sites. 

{\bf Qualitative considerations.} We first provide an intuitive description of the particle transport mechanism. In the strong disorder limit, $t\ll W$, typical hops of electrons between neighboring sites are off-resonant and therefore suppressed. However, the particles on the singly occupied sites have spin degree of freedom. Virtual hops between singly occupied sites generate an $SU(2)$ symmetric exchange interaction between their spins. The typical exchange constant ${\mathcal{J}}(\rho_s)$, estimated below, is suppressed in parameter $(t/W)$, and depends strongly on the density of {\it singlons} (singly occupied sites), ${\mathcal{J}}(\rho_s)\ll W$. Owing to the $SU(2)$ symmetry, according to Refs.~\cite{VasseurHotChains,PotterSymmetry,Abanin16, BarLev2016,ProtopopopovEtAl2018} the spin degrees of freedom delocalize and are expected to form a thermal bath. Further,  particle hopping processes couple to the spin bath: for example, there is a process of a particle hopping with a spin-flip, accompanied by flipping the spin on one of the neighboring singly occupied sites. The spin bath has a continuous spectrum, and can provide an energy mismatch to enable such a hopping process, leading to  the delocalization of particles. We note that such delocalization mechanism was discussed recently\cite{ParameswaranGopa} in the context of transport in a disordered, spin-incoherent Luttinger liquid.

The particle-number degree of freedom delocalizes, however, particle hopping processes are parametrically slow (the precise estimate is derived below). To understand the origin of the slow particle hopping rates, let us consider a simple initial state: all sites are singly occupied, with spins pointing in random directions. Let us also introduce one hole, and ask how quickly the hole would move.  The simplest process is that of the hole hopping to one of the nearest neighbor site. In this case, the typical energy mismatch is $\Delta E\sim W$, while exchange constant for the spin system is ${\mathcal J}_0\sim \frac{t^2 U}{W^2}\ll \Delta E$ (assuming limit of weak interactions, $U\ll W$). It has been shown \cite{Abanin15} that the narrow bandwidth of a thermal bath leads to parametrically long relaxation time scales, for processes with energy transfer much larger than the bath bandwidth. More precisely, the rate of the charge hopping process described above is given by: 
\begin{equation}\label{eq:estimate}
\Gamma\propto e^{-|\Delta E|/{\mathcal J_0}}\sim e^{-W^3/t^2 U}
\end{equation}
This illustrates why charge transport is slow in the limit of strong disorder. Similar to the variable-range hopping, one should consider processes where a particle hops between sites situated some distance away, and find the optimized (largest) hopping rate. Below we perform such an optimization, finding the radius of optimal hops. We find the corresponding hopping rate, which is faster than the above equation (\ref{eq:estimate}), derived for a nearest-neighbor hops predicts, but still parametrically slow.

\begin{figure}
\includegraphics[width=230pt]{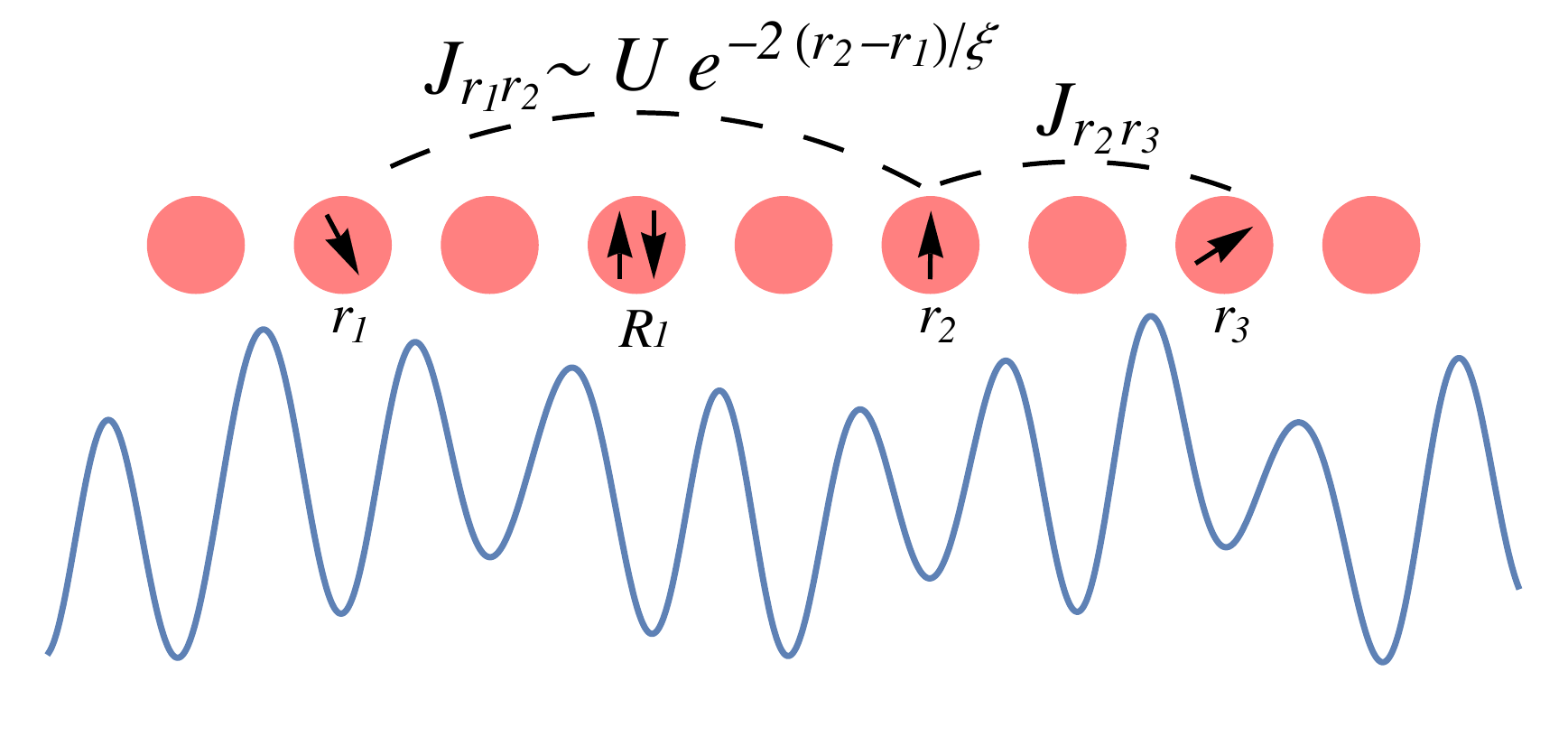}
\caption{A schematic of the disordered  Hubbard model. 
The solid line represents disorder potential, and the circles are lattice sites. Arrows represent spins of particles situated on some sites. The charge degrees of freedom are nearly localized for a parametrically long time. The state can be characterized by the positions of singlons ($r_i$) and doublons ($R_i$). Spin degrees of freedom are delocalized by the exchange interaction.
}
\label{Fig:Singlonsdublons}
\end{figure}

{\bf ${\it SU}(2)$ symmetry  and spin dynamics.} We start our analysis by estimating the exchange interaction between spins. As we expect the particle dynamics to be much slower than those of spins, we can first completely neglect the motion of particles and focus on the spin dynamics on singly occupied sites. We denote the singly-occupied sites by $r_i$ and the doubly occupied sites by $R_i$, see Fig. \ref{Fig:Singlonsdublons}, and their densities by $\rho_s$ and $\rho_d$, respectively.  The dynamical degrees of freedom are then the spins of the unpaired particles,
$ S^\alpha_{r_i}=c^+_{r_i\sigma}s^\alpha_{\sigma\sigma^\prime}c_{r_i\sigma^\prime}$.  Virtual particle hops give rise to an effective Hamiltonian for the spin degrees of freedom; the form of this Hamiltonian  is fully determined by the $SU(2)$ symmetry 
\begin{equation}
H_{\rm spin}=\sum_{\langle r_i, r_j\rangle} J_{r_i, r_j}{\bf S}_{r_i}{\bf S}_{r_j}+\ldots
\label{Eq:Hspin}
\end{equation}
Here, $\ldots$ denote the multi-spin interaction terms that are parametrically small at large disorder.

The coupling $J_{r_i, r_j}$ between two spins at distance $r_i-r_j$ arrises in the $2(r_i-r_j)$-th order of the perturbation theory in the hopping amplitude $t$. For two spins occupying adjacent sites,  we obtain
\begin{equation}
J_{r_i, r_{i+1}}=-\frac{4t^2 U}{(\epsilon_i-\epsilon_{i+1})^2-U^2}.
\label{Eq:Jii+1}
\end{equation}

The expressions for the couplings $J_{r_i, r_j}$  become especially simple in the limit of weak interaction, $U\ll W$.
Indeed, the Hamiltonian (\ref{Eq:H0}) can be rewritten in terms of  fermionic operators $a_i$  corresponding to the exact single-particle eigenstates $\psi_i$
\begin{equation}
H=\sum_{i}\tilde{\epsilon}_i a^{+}_{i\sigma}a_{i\sigma} +U\sum_{i, j, k, l} M_{ijkl}a^{+}_{i\uparrow} a_{j\uparrow} a^{+}_{k\downarrow} a_{l\downarrow}
\label{Eq:HTrans}
\end{equation} 
where the matrix elements $M_{ijkl}\equiv\sum_{\tilde{i}}\psi_i^*(\tilde{i})\psi_{j}(\tilde{i})\psi_k^*(\tilde{i})\psi_l(\tilde{i})$ decay exponentially (with the localization length $\xi\sim1/\ln(W/t))\ll 1$)  as functions of all the four distances $|i-j|$, $|j-l|$, etc. 
 The matrix elements $M_{ijkl}$ with two pairs of coinciding indices, $M_{ijji}$, are of special interest to us, because they determine the exchange couplings between spins $i,j$ in Eq. (\ref{Eq:Hspin}):
\begin{equation}
J_{r_i, r_j}\sim U M_{ijji}\,, \qquad J_{\rm typ} \sim Ue^{-2/\xi \rho_s}.
\label{Eq:J}
\end{equation}
The couplings $J_{r_i, r_j}$ given by (\ref{Eq:J}) are insensitive to the presence of doublons in the system. This feature 
is  preserved also beyond the limit $U\ll W$ as long as the condition $t\ll W$ is fulfilled.

Equations (\ref{Eq:Hspin}) and (\ref{Eq:J})  describe the dynamics of unpaired particle spins, the most mobile degrees of freedom in the system, in terms of a random Heisenberg model. It was recently argued~\cite{PotterSymmetry, Abanin16} that, due to $SU(2)$ symmetry leading to proliferation of long-range resonances, this model remains delocalized even in the case of relatively strong exchange-coupling disorder (which arrises naturally in our case at $\rho_s\ll 1$, due to the broad distribution of the localized wave function amplitudes). Therefore, the particle spins in our system are expected to form a bath with continuous spectrum characterized by a spectral function
\begin{equation}
f^{r_i r_i^\prime}_{\rm spin}(\omega)=\int dt \langle S_{r_i}^+(t)S_{r_{i^\prime}}^-(0)\rangle e^{-i\omega t}.
\label{Eq:fDef}
\end{equation}
The properties of the spin bath are  controlled by the typical exchange coupling $J_{\rm typ}$.  As we consider a random initial state with high energy density   the averaging in Eq. (\ref{Eq:fDef}) is effectively over infinite temperature ensemble.  We expect 
$f^{r_i r_j}_{\rm spin}(\omega)$ to decay fast with distance $r_i-r_i^\prime$ and focus on its fully  local 
limit $f_{\rm spin}(\omega)\equiv f^{r_ir_i}_{\rm spin}(\omega)$. 
The frequency dependence of the spectral  function $f_{\rm spin}(\omega)$  in disordered spin systems can be   rather complicated \cite{Serbyn-16}.
However, as the atom hops between localized  states typically involve energy mismatch $\omega\gg J_{\rm typ}$ only its  high-frequency asymptotic behavior given by ~\cite{Abanin15, AbaninDeRoeckPRB2017}
\begin{equation}
f_{\rm spin}(\omega)\sim 
\frac{1}{J_{\rm typ}} e^{-{\mathcal C}|\omega|/J_{\rm typ}}, \qquad \omega\gtrsim J_{\rm typ}.
\label{Eq:f}
\end{equation}
is relevant for our purposes. In this equation, ${\mathcal C}$ is a non-universal constant of order one, which we will take to be one for simplicity. 
The exponential decay of the spectral function at large frequency $\omega\gg J_{\rm typ}$ is a generic phenomenon which arises due to the fact that in order to absorb/emit energy $\omega\gg J_{\rm typ}$, a large number of spins $N\sim |\omega|/J_{\rm typ}$  has to be rearranged.

{\bf Spin bath and particle dynamics. } The interaction term in Eq. (\ref{Eq:HTrans}) contains matrix elements $M_{ijki}$ describing particle hops from site $j$ to site $k$ assisted by spin flip at site $r_i$,
\begin{equation}
H_{{\rm sc}}=\sum_{i, j, k} {\cal J}_{ijk}\left(S_{r_i}^+ a^+_{k\downarrow}a_{j\uparrow}+{\rm h.c.} \right).
\label{Eq:HSpCh}
\end{equation}
where ${\cal J}_{ijk}\sim U \exp\left[-\max(|j-k|, |r_i-k|, |r_i-j|)/\xi\right]$. (In addition, the interaction term in Eq.(\ref{Eq:HTrans}) contains matrix elements responsible for the spin exchange and matrix elements of the type $M_{iiik}$ that renormalize the single-particle hopping amplitudes for doubly-occupied sites).

\begin{figure}
\includegraphics[width=1\columnwidth]{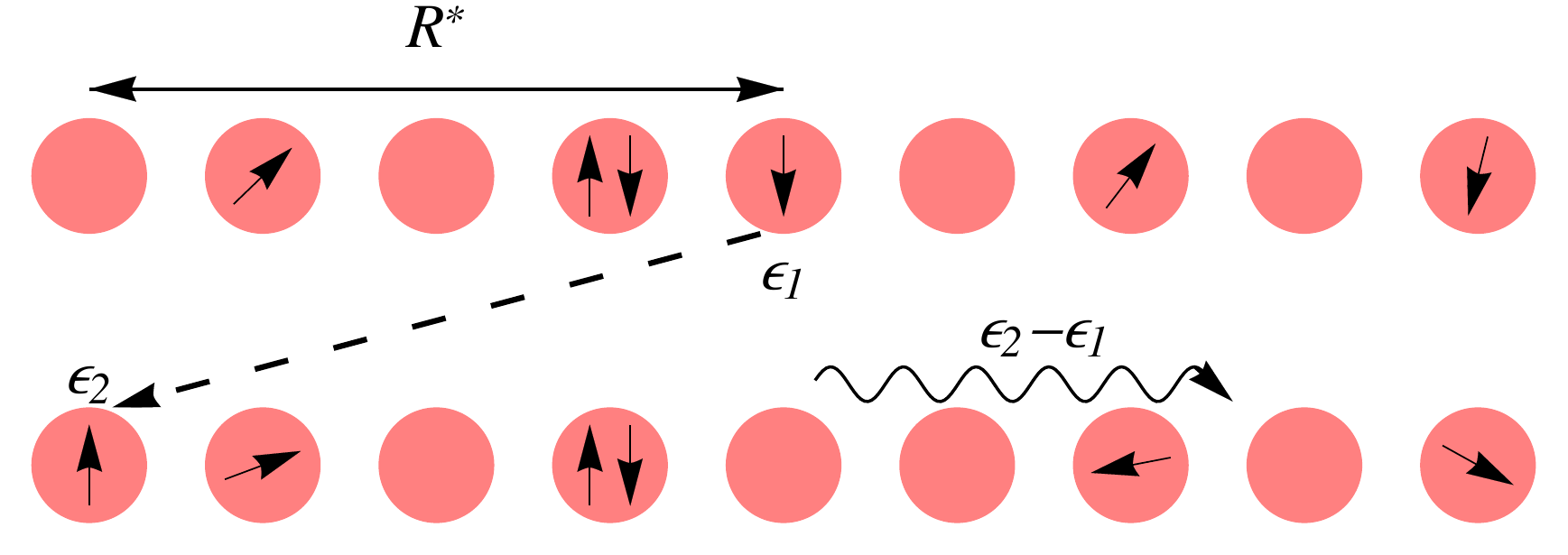}
\caption{Illustration of a particle hop assisted by an excitation of a spin bath: a particle can hop between sites 1,2, and the mismatch energy is provided by the spin bath. There is an optimal hopping distance $R^*$, which depends on the density of the singlons and doublons, see Eqs.(\ref{Eq:JtypStrong},\ref{Eq:GammaMain}).}
\label{Fig:Hopping.eps}
\end{figure}

Typically strongly off-resonant, the processes described by Eq. (\ref{Eq:HSpCh})  are very slow. Thus, we can treat the fermionic and spin operators in Eq. (\ref{Eq:HSpCh}) as describing independent degrees of freedom ( a kind of spin-charge separation) and consider the dynamics of a single particle in the environment of the  spin bath. Spin  bath  leads then to particle number dynamics via a mechanism reminiscent of the variable range hopping (VRH) in semiconductors.
 Specifically,  the particle transition rate from site $j$ to site $k$ (see Fig. \ref{Fig:Hopping.eps}) is given by the Fermi golden rule 
 as
 \begin{equation}
 \Gamma_{j\rightarrow k}\sim\sum_{r_i}{\cal J}^{2}_{ijk}f_{\rm spin}(\epsilon_i-\epsilon_k)
 \end{equation}
 where we have taken into account the short-range nature of correlations in the spin bath. A particle hop by distance $R$ involves energy mismatch $|\epsilon_i-\epsilon_k|\sim W/R(1-\rho_d)$ (with $R(1-\rho_d)$ being the number of available final states within distance $R$) that can be compensated by an excitation of the spin  bath. Using the bath spectral function (\ref{Eq:f}) and anticipating that typical $\omega\gg J_{\rm typ}$ we find the rate of such a process:
\begin{equation}
\Gamma(R)\sim\frac{U^2 R \rho_s}{J_{\rm typ}} e^{-2 R/\xi}e^{-W/R(1-\rho_d)J_{\rm typ}}
\label{Eq:GammaBasic}
\end{equation}
where the prefactor originates from the summation over the coordinate of the spin involved in the process. 

The rate (\ref{Eq:GammaBasic}) should be optimized with respect to the hopping distance $R$ leading to
\begin{equation}
R^*\sim \sqrt\frac{W\xi}{2(1-\rho_d)J_{\rm typ}}\,, \qquad \Gamma(\rho_s) \sim 
\frac{U^2 R^* \rho_s}{J_{\rm typ}} e^{-4 R^*/\xi}.
\label{Eq:GammaMain}
\end{equation}
According to Eqs. (\ref{Eq:GammaMain}) and (\ref{Eq:J}) $R^*\gg 1/\rho_s$ and the charge transport involves hops much longer  than the  average inter-particle distance. 

Eq.(\ref{Eq:GammaMain}) shows that the delocalization of spin degrees of freedom in the Hubbard model leads to a finite but exponentially slow particle number relaxation via variable-range hopping. We stress that, in contrast to the conventional variable-range hopping in semiconductors mediated by phonons and controlled by the temperature, our transport channel is mediated by spin excitations, occurs at infinite temperature and is controlled by disorder.  It follows from Eq.(\ref{Eq:J}) for $J_{\rm typ}$ that the relaxation rate depends strongly on the density of free spins in the system and is maximal for $\rho_s\lesssim 1$ (we assume for simplicity that $\rho_d=0$)
\begin{equation}
\Gamma(\rho_s=1)\sim \frac{U W^2}{t^2}\exp\left[-\sqrt{\frac{8 W^3}{Ut^2}\ln \frac{W}{t}}\right].
\end{equation}
The particle hopping will lead to thermalization and decay of the initial CDW patterns. However, we emphasize the very strong (doubly exponential) dependence of the particle hopping rate on the density of singly occupied sites $\rho_s$, which follows from Eq.(\ref{Eq:GammaBasic}). Therefore, initial states with low density of singlons will appear fully localized for any reasonable time of observation. 

\begin{figure}
\includegraphics[width=250pt]{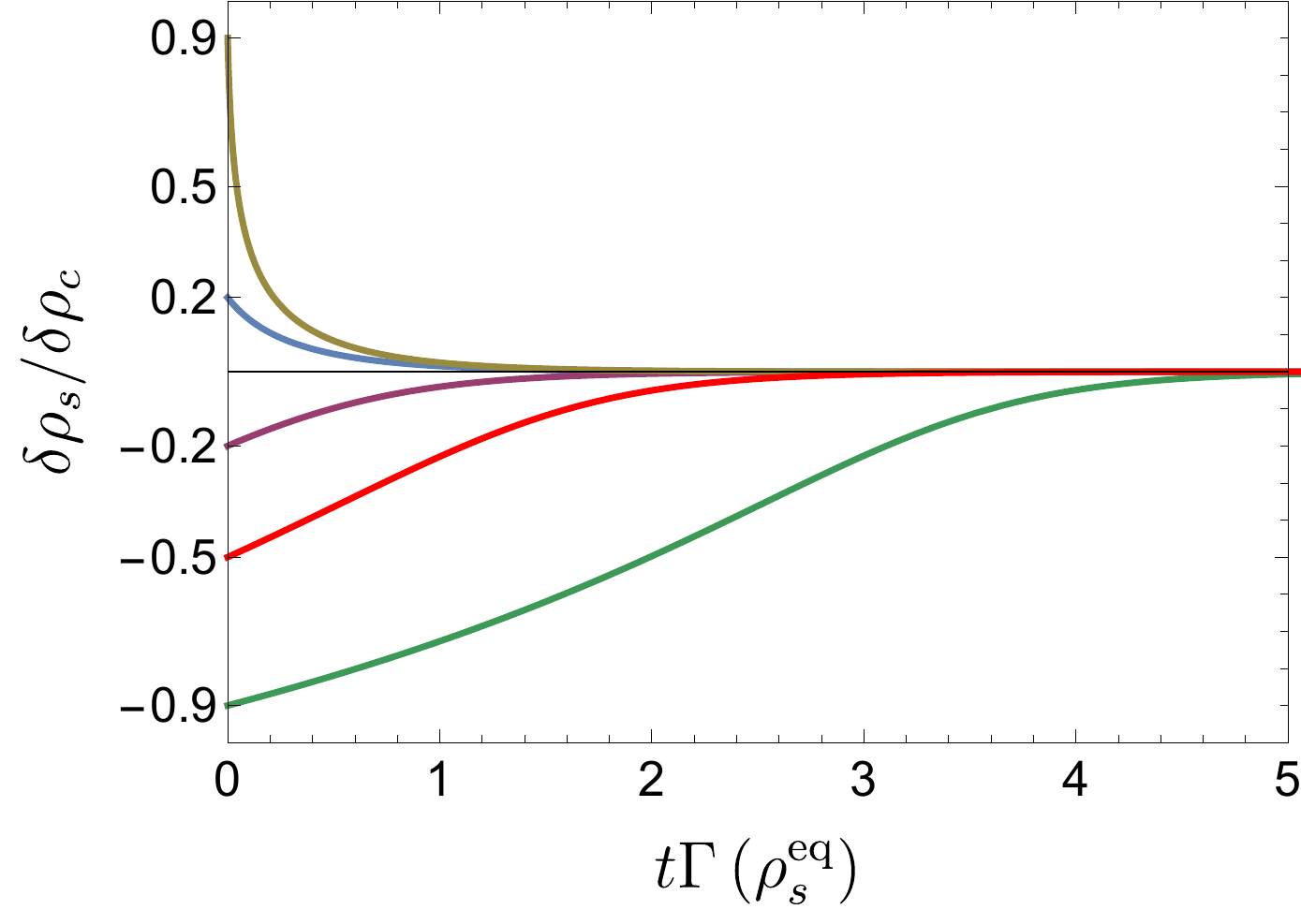}
\caption{Relaxation of the density of singlons to equilibrium value as  described by Eq. (\ref{Eq:RateEq}).
Vertical  axis shows the ratio $\delta\rho_s/\delta \rho_c$ with the characteristic density scale given by Eq. (\ref{Eq:deltarhoc}) for various values of the initial deviation $\delta \rho_s$. The parameters used to generate the plot  are: 
$W=1$, $\xi=0.5$, $U=0.1$, $\rho=0.5$. 
 The time is measured in units of equilibrium relaxation time, $1/\Gamma(\rho_s^{\rm eq})$. While the relaxation of positive $\delta \rho_s$ occurs on the time scale $1/\Gamma(\rho_s^{\rm eq})$,  large negative $\delta \rho_s$ persist till much longer times set by $\Gamma[\rho_s(t=0)]$.  }
\label{Fig:SinglonRelaxation}
\end{figure}

The same strong dependence of $\Gamma$ on $\rho_s$ will manifest itself  in a strongly non-exponential and asymmetric relaxation of the density of singlons to its equilibrium value $\rho_s^{\rm eq}$ which, in the small $U$ limit, is dictated by the overall density of particles
\begin{equation}
\rho_s^{\rm eq}=\rho-\frac{\rho^2}{2}.
\end{equation} 
Indeed, we can model this relation by a simple rate equation 
\begin{equation}
\frac{d \rho_s}{dt}=\frac{4\Gamma(\rho_s)(\rho_s^{\rm eq}-\rho_s)}{2-\rho+\rho_s}.
\label{Eq:RateEq}
\end{equation}
It follows now that exponentially small deviations of $\rho_s$ from equilibrium density (we assume for simplicity the low-density limit $\rho\ll 1$) 
\begin{equation}
|\delta\rho_s| \equiv|\rho_s-\rho_s^{\rm eq}| \ll\delta\rho_c\equiv \sqrt{\frac{U \xi^3\rho^4}{W}}e^{-\frac{1}{\xi \rho}}
\label{Eq:deltarhoc}
\end{equation}
follow straight exponential-in-time relation  with the time scale set by $\Gamma(\rho)$. 
The same time scale effectively controls the relaxation of larger {\it positive} deviations of $\rho_s$ which consists now of a rapid decrease of $\delta\rho_s$ to $\delta\rho_c$ followed by exponential relaxation.  
 On the other hand,  for larger {\it negative} deviations $\delta\rho_s<-\delta\rho_c$ the initial state is the bottle neck in the relation process and the characteristic time is set by the the initial density of doublons.  The evolution of singlon density for various initial conditions  is  illustrated in Fig.~\ref{Fig:SinglonRelaxation}. 
 
{\bf Strong interaction.} Much of the analysis presented above can be extended to the case of strong interaction $U\gg W$ (but still in the strong localization limit $t\ll W$). Straightforward power counting shows that in this regime typical exchange coupling obeys  [cf. Eq. (\ref{Eq:Jii+1})]
\begin{equation}
J_{\rm typ}\sim \frac{W^2}{U}e^{-2/\xi\rho_s}.
\label{Eq:JtypStrong}
\end{equation} 
Further, the  matrix element for the spin bath assisted hopping, ${\cal J}_{i jk}$ is of the form 
\begin{equation}
{\cal J}_{ijk}\sim \frac{W^2}{U} \exp\left[-|j-k|/\xi\right].
\end{equation}
Important difference between weak and strong interaction limits comes when counting the number of final states available for a particle hop. One needs now to distinguish between single particle hops [typical energy mismatch $W/R(1-\rho_d-\rho_s)$], "doublon hops" [a process where  a particle hops from a site occupied by doublon  to form another  doublon leaving behind an unpaired electron, typical energy mismatch $W/R \rho_s$] and doublon decay into unpaired spins [energy mismatch $U\gg W$]. 

Among these processes, only the last one leads to  the equilibration between singlon and doublon densities. It is also the slowest one  of the three, because it has to involve {\it nearest-neighbor}  hopping.  It is characterized by the rate [cf. Eqs. (\ref{Eq:JtypStrong}) and (\ref{Eq:f})]:
\begin{equation}
\Gamma_{sd}\propto\exp\left[-\frac{U^2 e^{2/\xi\rho_s}}{W^2}\right].
\label{GammaSD}
\end{equation}

In contrast, the singlon and "doublon" hopping processes are of the variable-range type.  In full analogy with Eq. (\ref{Eq:GammaMain}), we find the corresponding rates 
\begin{equation}
\Gamma_{s (d)}\propto \exp\left[-4\sqrt{\frac{U}{2W\alpha_{s(d)}}}e^{1/\xi\rho_s} \right].
\label{GammaSandD}
\end{equation}
where $\alpha_s=1-\rho_d-\rho_s$ and $\alpha_d=\rho_s$.  For moderate lattice filling $\rho\sim 1$ the rates $\Gamma_s$ and $\Gamma_d$ are comparable (in log scale). On the other hand, in the low-density limit $\Gamma_d\ll \Gamma_s$ and doublons are practically frozen. 

 It is interesting to apply the above results to the initial state where doublons are positioned on odd sites, while even sites are empty (charge-density-wave state). Such an initial state does not have single occupancies, and therefore the  spin bath cannot form.  The analysis presented above  suggests then that such a state has diverging relaxation times [see Eqs. (\ref{Eq:GammaMain}), (\ref{GammaSD}) and (\ref{GammaSandD})].  We stress that for repulsive interaction $U$ this means the existence of a non-thermalizing sector in the Hilbert space (with exponentially many states) in the middle of the many-body energy band. Detailed analysis of dynamics in the vicinity of this sector is an interesting direction for future work. 

{\bf Symmetry breaking and MBL.}  The delocalization of spin and, ultimately, of the particle number degrees of freedom in our system, 
rely on the $SU(2)$ symmetry of the Hamiltonian (\ref{Eq:Hspin}) . When $SU(2)$ symmetry is broken (e.g. by random magnetic filed), the strongly-disordered spins remain localized and so do the particles. As pointed out in Ref.~\cite{Znidaric16}, it is not enough to break the $SU(2)$ symmetry by application of a  {\it uniform} magnetic filed as it would only couple to the $z$-projection of the total spin of the system, which is an exact integral of motion; thus, the many-body eigenstates will not be modified. However, we expect a uniform {\it gradient} of magnetic field (that is easy to realize in experiment) to suffice for triggering MBL. Note that already a very weak gradient causing Zeeman splitting of the order of $J_{\rm typ}$ between nearest-neighbor spins is sufficient. A detailed study of the field-induced transition to MBL states is an interesting direction for future research.  
 
{\bf Conclusions.} We have studied equilibration and particle transport in strongly disordered Fermi-Hubbard model. We have shown that 
$SU(2)$ symmetry of the Hamiltonian precludes localization and  eventually leads to exponentially slow particle transport, Eqs. (\ref{Eq:GammaMain}), (\ref{GammaSandD}) and (\ref{GammaSD}). The transport  mechanism is reminiscent of the variable-range hopping, but it is mediated by spin degrees of freedom.  
Breaking $SU(2)$ symmetry by a weak magnetic field gradient can induce transition to an MBL state. 
Our predictions can be tested  in a quench experiment  with ultracold atoms.  Due to the strong dependence of the particle-number relaxation rate on the density of singlons, the preferable initial state would the one with a high density of singlons, because this would give rise to the fastest particle dynamics. One possibility would be to prepare an initial state where majority of sites are singly occupied (and spins are initially random), and there is a small density of holes, dynamics of which will be monitored. 

{\bf Acknowledgements.} We thank Wen Wei Ho, Uli Schneider,  Immanuel Bloch, Antonello  Scardicchio and Eugene Demler for illuminating discussions. This work was supported by the Swiss National Science Foundation and  by Russian Science Foundation under Grant No. 14-42-00044. 

\bibliography{}

\end{document}